\documentclass[twocolumn,tighten,trackchanges]{aastex631}
\hypersetup{linkcolor=blue,citecolor=blue,filecolor=cyan,urlcolor=blue}
\shorttitle{Solar-cycle Variations in the Near-Surafce Shear Layer}
\shortauthors{Sen et al. }

\begin{document}

\title{Solar-cycle variations in meridional flows and rotational shear within the Sun's near-surface shear layer}

\author[0000-0003-2694-3288]{Anisha Sen} 
\email{anisha.sen@iiap.res.in}
\affiliation{Indian Institute of Astrophysics, II Block Koramangala, Bengaluru 560 034, India}
\affiliation{Pondicherry University, R.V.Nagar, Kalapet, Puducherry 605014, India}
\author[0000-0003-0003-4561]{S.P. Rajaguru}
\email{rajaguru@iiap.res.in}
\affiliation{Indian Institute of Astrophysics, II Block Koramangala, Bengaluru 560 034, India}
\author[0000-0001-8042-2358]{Abhinav Govindan Iyer}
\affiliation{Indian Institute of Astrophysics, II Block Koramangala, Bengaluru 560 034, India}
\author[0000-0002-2632-130X]{Ruizhu Chen}
\affiliation{Department of Physics, Stanford University, Stanford, CA 94305-4060, USA}
\affiliation{W. W. Hansen Experimental Physics Laboratory, Stanford University, Stanford, CA 94305-4085, USA}
\author[0000-0002-6308-872X]{Junwei Zhao}
\affiliation{W. W. Hansen Experimental Physics Laboratory, Stanford University, Stanford, CA 94305-4085, USA}
\author[0000-0003-1860-3697]{Shukur Kholikov}
\affiliation{National Solar Observatory, Boulder CO, USA}
\affiliation{National Research University TIIAME, Kori Niyoziy 39, Tashkent 100000, Uzbekistan}




\begin{abstract}

Using solar-cycle long helioseismic measurements of meridional and zonal flows in the near-surface shear layer (NSSL) of the Sun, we study their
spatio-temporal variations and connections to active regions. We find that near-surface inflows towards active latitudes are part of
a local circulation with an outflow away from them at depths around 0.97 R$_\odot$, which is also the location where the
deviations in the radial gradient of rotation change sign. These results, together with opposite signed changes, over latitude
and depth, in the above quantities observed during solar minimum period, point to the action of Coriolis force on large-scale flows 
as primary cause of changes in rotation gradient within the NSSL. We also find that such Coriolis force mediated changes in near-surface
flows towards active latitudes only marginally change the amplitude of zonal flow and hence are not likely its driving force. 
Our measurements typically achieve a high signal-to-noise ratio ($>$5$\sigma$) for near-surface flows but can drop to 3$\sigma$ near the 
base (0.95 R$_\odot$) of the NSSL. Close agreements between the depth profiles of changes in rotation gradient and in meridional flows measured 
from quite different global and local helioseismic techniques, respectively, show that the results are not dependent on the analysis techniques. 

\end{abstract}

\keywords{The Sun; Solar Cycle; Solar Activity; Solar Oscillations; Solar Rotation; 
Meridional and Zonal Flows; Helioseismology}


\section{Introduction} 
\label{sec:intro}

The near-surface shear layer (NSSL) \citep{thompson1996differential} of the Sun, situated just below the solar surface 
over depths to about 35 Mm, is marked by a swift increase in the rotation rate as depth increases \citep{schou1998helioseismic}.
Several helioseismic studies \citep{2002SoPh..205..211C,2010ApJ...720..494A,2014A&A...570L..12B,barekat2016solar,antia2022changes} have 
since established its overall structure and the later ones have also uncovered changes, over space and time, that
relate to active region magnetic fields and solar cycle. These latter solar cycle related changes connect the NSSL to the already 
well studied zonal flows or torsional oscillations that extend almost throughout the convection zone 
\citep{2001ApJ...559L..67A,2002Sci...296..101V}.
The structure and dynamics of the NSSL uncovered by helioseismology so far \citep{barekat2016solar,antia2022changes} pose
challenges to theoretical and simulation studies \citep{2000SoPh..192...59M,2016AstL...42..339K,2019ApJ...871..217M,2023AstL...49..754K}, 
and have also served to emphasise it as an important region where complex magnetohydrodynamic processes controlling the magnetic 
activity of the Sun play out. These latter aspects have attracted attention of modellers in exploring the roles of NSSL in 
dynamo processes \citep{2002ApJ...575L..41D,2005ApJ...625..539B,2011ApJ...727L..45P,2016ApJ...832...94K}. 
 
Although the zonal and meridional flows extend throughout the convection zone straddling the NSSL,
probing solar cycle related changes in them within it is expected to give clues to understanding not only
the origin of these flows but also of the NSSL and its maintenance \citep{2016AstL...42..339K,2019ApJ...871..217M}.
In this connection, understanding the dynamics of large-scale inflows towards active regions \citep{2001IAUS..203..189G,2004SoPh..220..371H,
hindman2009subsurface,2019ApJ...873...94B} and their cumulative contributions to the changes in zonal \citep{antia2022changes} and 
meridional flows \citep{2022A&A...664A.189P,mahajan2023removal} is important. Given that the Coriolis force plays central roles 
in the redistribution of angular momentum over depth \citep{lebedinsky1941rotation,2016AstL...42..339K}, it will be interesting
to examine how such forces on flows around active regions contribute to the changes in the NSSL.  
In fact, \cite{spruit2003origin} presented a model for the zonal flows in terms of geostrophic flows driven by sub-surface temperature 
variations caused by the increased radiative losses through the plages and small-scale magnetic fields surrounding sunspots. 
Helioseismic measurements, using ring diagram technique, showed that sunspots and surrounding plages drive 
a mean inflow with speeds of 20 to 30 ms$^{-1}$ and cyclonic circulation of 5 ms$^{-1}$ around their edges \citep{hindman2009subsurface}. 
The overall contributions of such flows to the variations in meridional flows have been assessed by \citet{2022A&A...664A.189P} and
\citet{mahajan2023removal}. The former authors have concluded that there remain signifcant solar cycle
variations in the meridional flows near the surface even after removing active region inflows.
For solar cycle 24 and early part of cycle 25, \cite{mahajan2023removal} deduced that the variations in the
background meridional flow are largely due to active regions, whereas the torsional oscillation remains an universal
phenomenon, as its magnitude or phase shows no significant alteration when active region neighborhoods are excluded. 
\cite{antia2022changes} have shown that the radial gradient of solar rotation rate is influenced by the solar 
cycle, exhibiting more significant changes in active latitudes than in neighboring higher latitudes. 

In this paper, through a joint analysis of helioseismic measurements of meridional and zonal flows, we explore the 
depth structure of changes due to flows around active regions and examine how they influence the gradient of solar rotation and its 
variation over depth and time within the NSSL. 
The rest of the paper is structured as follows: Section \ref{sec: data} outlines the data utilized, followed by a 
description of the analysis technique in Section \ref{sec: analysis}. Section \ref{sec: result} presents our findings, 
and Section \ref{sec: conclusion} discusses the implications of our results.
 
\section{Data} \label{sec: data}

We use helioseismic data from the space-borne Helioseismic and Magnetic Imager
(HMI) \citep{scherrer2012helioseismic} onboard NASA's Solar Dynamics Observatory(SDO), which has been observing the Sun
since March 2010. For seismic inversions of solar interior rotation rate and changes (zonal flows) in it, we have used 
oscillation frequencies and their splittings measured from the HMI 72-day time-series of spherical harmonic coefficients of 
global oscillation modes (series hmi.V\_sht\_modes) \citep{antia2022changes}. The 72-day datasets we have used start from 
set number 6328 with date 2010.04.30 and end at set number 10720 with date 2022.07.22. 

For time-distance helioseismic measurements of meridional flow, we use the HMI/SDO Dopplergrams with a 
45-sec cadence, spanning 13 years from May 2010 to April 2023. Each day's data are tracked and Postel remapped to account for solar rotation 
and projection effects using the Stanford Joint Science Operations Center (JSOC) helioseismology pipeline, following 
the procedures decribed by \citet{zhao2013detection}.
Additionally, we utilize identically processed Global Oscillation Network Group (GONG) data for
the same meridional flow measurements. 
The network-merged data, obtained from GONG stations around the globe over the same time period, is used in this study. 
The final data products used for travel time measurements, from both HMI and GONG, have a binned down spatial resolution 
of 0.36 deg/pix.

We also employ local time-distance helioseismic inversions for horizontal
velocity fields \citep{2012SoPh..275..375Z}, publicly available in the JSOC time-distance helioseismology pipeline{\footnote{\textcolor{blue}
{http://jsoc.stanford.edu/data/timed/}}}, along with the HMI line-of-sight magnetograms, to check 
and establish connections between flows around individual active regions and our measurements of global-scale changes in 
meridional and zonal flows. For comparing time - latitude profiles of flows with that of sunspots,
we use data on the latter from the National Oceanic and Atmospheric Administration’s Solar Region Summary.

\section{Analysis Techniques} \label{sec: analysis}

\subsection{Meridional Flow}
\label{subsec: meridional}

We measure travel times of acoustic waves ($p$ modes) that propagate between two surface locations,
connected by its path through the interior, using the technique of time-distance helioseismology \citep{duvall1993time}. 
The measurement geometry involves North-South (N-S) and West-East (W-E) oriented arcs (36$^\circ$ wide) for measuring the meridional flows and 
the center-to-limb systematics (CLS) \citep{zhao2012systematic}, respectively, employing the ``point-to-point'' cross-correlation of the Doppler signals 
as explained in \citet{rajaguru2015meridional}. The CLS in travel times are removed following the same procedure as originally 
suggested by \citet{zhao2012systematic} and adopted in almost all time-distance measurements of meridional circulation to date.
Both the HMI and GONG datasets are processed and subjected to exactly the same method of analysis, and the measurements used here are
part of an ongoing detailed study \citep{rzchenetal2025} to understand the additional systematic differences between GONG 
and HMI travel times reported by \citet{2020Sci...368.1469G}. These additional systematics affect inferences 
only of deeper (below 0.9$R_{\odot}$) layers and also change slowly over solar cycle timescales. 
Since we are studying changes in meridional flows by subtracting a long 
time average (over the solar cycle) from each 1-year running average solution obtained at each month, 
much of the systematics get subtracted out too. Additionally, to take care of the
surface magnetic effect in flow measurements \citep{2015ApJ...805..165L,2017ApJ...849..144C}, we mask out
active regions in input Doppler data that exceed a threshold of 40 G in the 0.36 deg/pix resolution HMI LOS magnetograms.
Note that we recover meridional flows from the inversions for the stream function, which satisfies the continuity equation and thus the mass 
conservation constraint is built into the inversion scheme \citep{rajaguru2015meridional}.

\subsection{Zonal Flow}
\label{subsec: zonal}

Solar interior rotation rate as a function of radius ($r$), latitude ($\theta$), and time ($t$), 
$\Omega$ = $\Omega(r,\theta,t)$, is determined through 
seismic inversions of oscillation frequency splittings (described in Section \ref{sec: data}) by 
adapting the 2D Regularized Least Squares (RLS) technique \citep{antia1998solar}. 
The rotation rate is represented as a product of cubic B-splines in solar radius and cosine of the colatitude.
Uniformly spaced 20 knots or nodes in the cosine of colatitude, and 50 knots in the acoustic radius have been used for 
inversion. Since the main goal is to study changes in the NSSL, we calculate residuals by subtracting solar cycle-long 
averages. Our datasets covered Cycle 24 and initial part of cycle 25, however we subtract only Cycle 24 average 
from the individual measurements. Thus the residual rotation rate is given by, $\delta\Omega(r,\theta,t) = \Omega - <\Omega>_{C24}(r,\theta)$
and the residual of the dimensionless radial gradient of rotation rate \citep{antia2022changes}{\footnote{Rotation inversions used
here are the same as published by \citet{antia2022changes} with addition of about one year longer data sets, and were provided
by H.M. Antia.} (residual rotational shear, hereafter) is,
$\delta(\partial~{\rm{log}}~\Omega/\partial~{\rm{log}}~r)(r,\theta,t) = (\partial~{\rm{log}}~\Omega/\partial~{\rm{log}}~r) - 
<(\partial~{\rm{log}}~\Omega/\partial~{\rm{log}}~r)>_{C24}$, where $< >_{C24}$ stands for time average over Cycle 24. 
We apply a 1-year running mean to smooth out the fluctuations in the measurements at each $r$ and $\theta$.

\begin{figure*}[!htbp]
\gridline{\fig{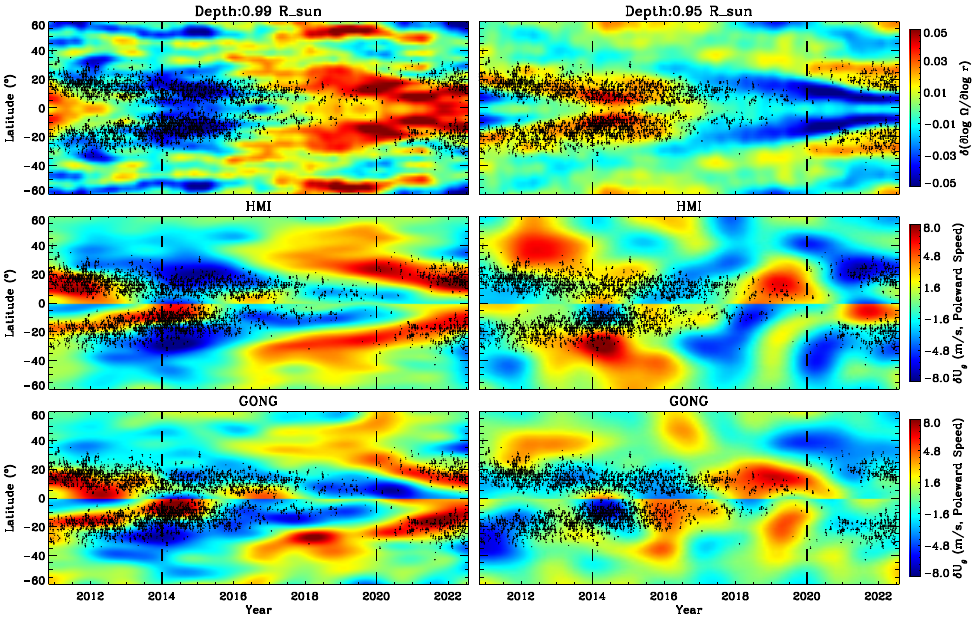}{1\textwidth}{}          } 
\caption{Changes (Cycle 24 subtracted) in dimensionless radial gradient of rotation rate, $\delta(\partial~{\rm{log}}~\Omega/\partial~{\rm{log}}~r)$,
from HMI global helioseismic rotation inversions (top panels) and those in the meridional flow, $\delta U_{\theta}$, (middle and bottom panels 
from HMI and GONG, respectively) as a function of latitude and time at 0.99 $R_{\odot}$ (left) and at 0.95 $R_{\odot}$ (right). 
Sunspot locations are overplotted as black dots. The two vertical dashed lines mark Cycle 24 maximum (2014) and minimum (2020).}
\label{fig1}
\end{figure*}

\begin{figure}[!htbp]
\includegraphics[width=0.5\textwidth]{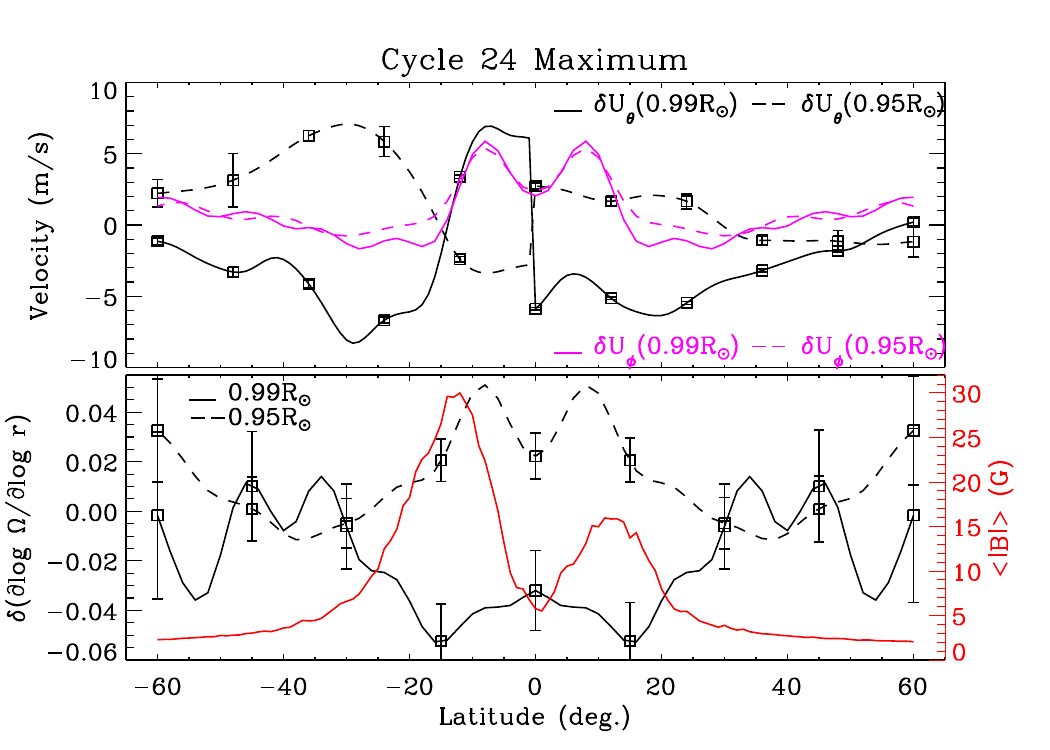}
\caption{Latitude profiles of changes in meridional flow ($\delta U_{\theta}$, in black) and zonal flow ($\delta U_{\phi}$, in pink), averaged
over one year around the Cycle 24 maximum, for depths 0.99 $R_{\odot}$ ({\it solid}) and 0.95 $R_{\odot}$ ({\it dashed}) in the top panel.
The same in the radial gradient of rotation, $\delta(\partial~{\rm{log}}~\Omega / \partial~{\rm{log}}~r)$, along with the longitudinally
averaged unsigned magnetic field (right y-axis), is shown in the lower panel. The errorbars shown represent errors estimated in the inversion method
see Section \ref{subsec: results_1} for details.}
\label{fig2}
\end{figure}

\begin{figure*}[!htbp]
\gridline{\fig{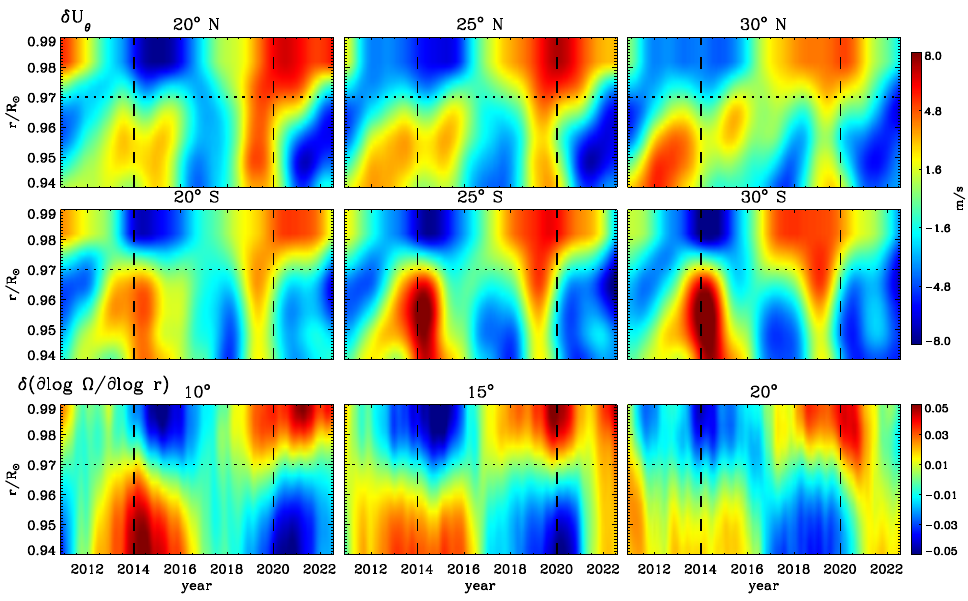}{1\textwidth}{}} 
\caption{Connections between the time vs depth profiles of changes in meridional flow ($\delta U_{\theta}$) and 
that in the radial gradient of rotation [$\delta(\partial~{\rm{log}}~\Omega / \partial~{\rm{log}}~r)$]. The top and middle panels show
$\delta U_{\theta}$ for the northern and southern latitudes (20$^{o}$,25$^{o}$ and 30$^{o}$), respectively. 
The north - south symmetric component of changes in $\delta(\partial log \Omega / \partial log r)$ are shown in the bottom
row for latitudes (10$^{o}$, 15$^{o}$, and 20$^{o}$). The two vertical dashed lines in each panel mark Cycle 24 maximum (2014) and minimum (2020),
and the horizontal dotted lines mark the depth 0.97 $R_{\odot}$. All the results here are from the HMI data.}
\label{fig3}
\end{figure*}

\begin{figure*}[t!]
\includegraphics[width=0.5\textwidth]{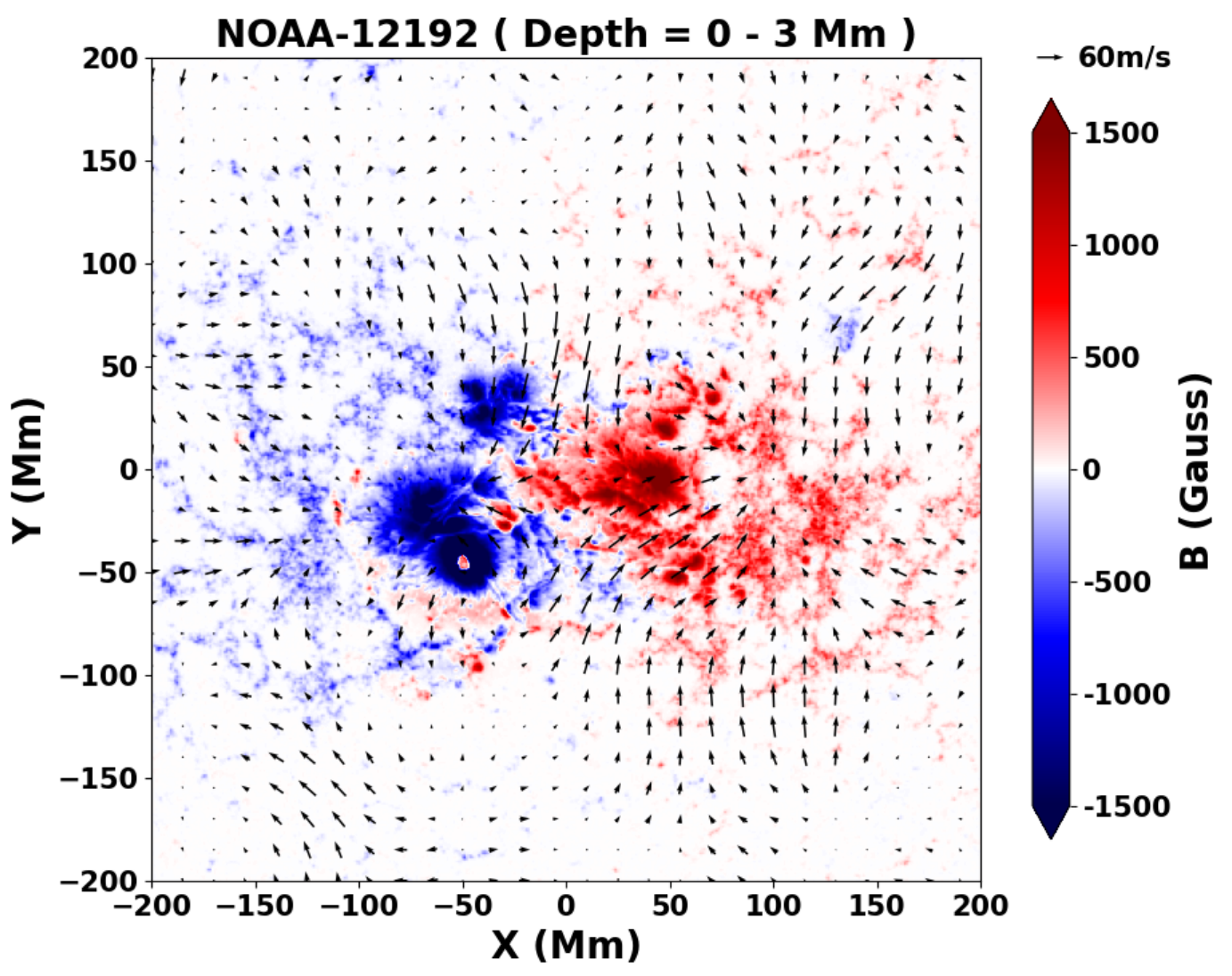}
\includegraphics[width=0.5\textwidth]{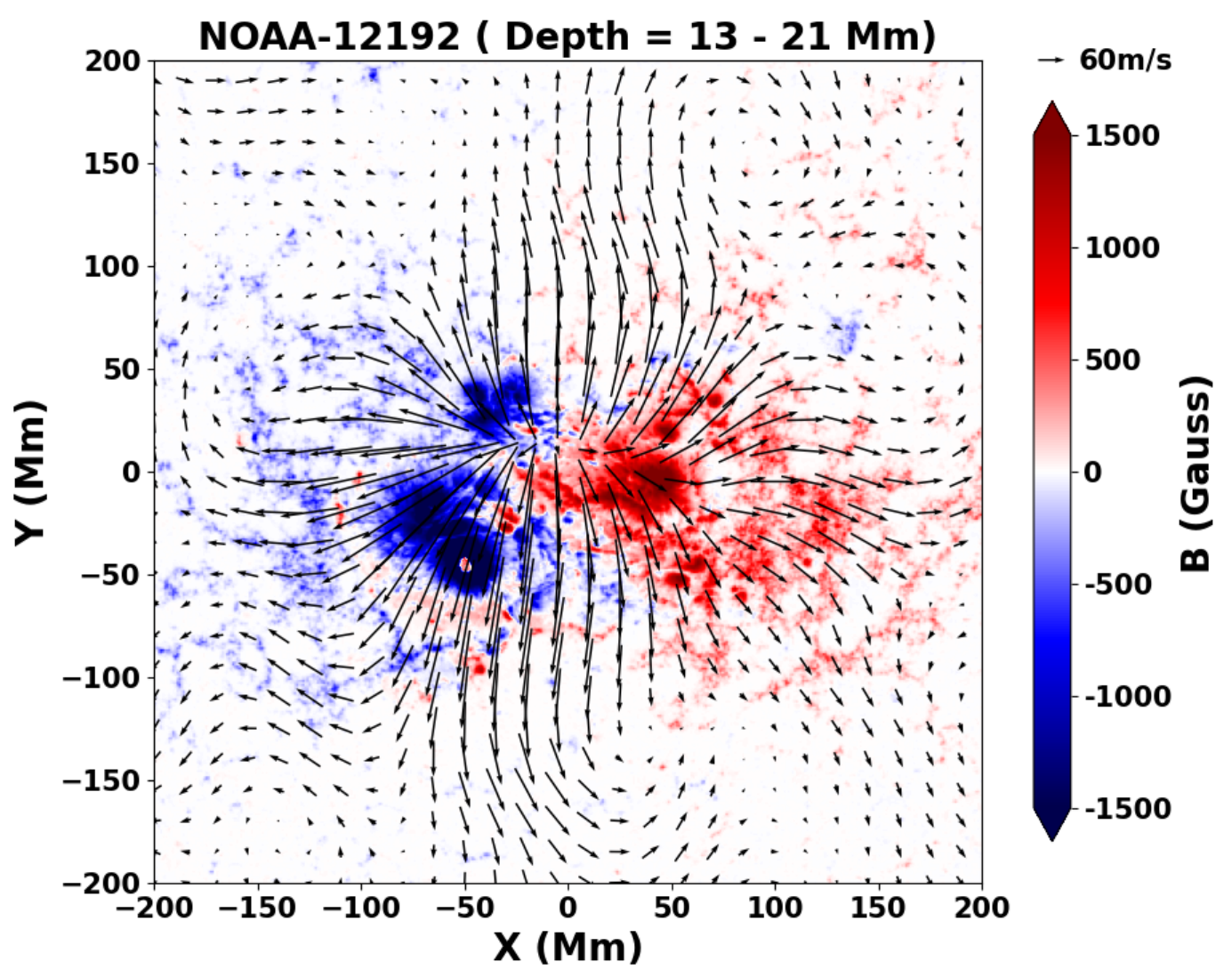}
\caption{Local time-distance helioseismic inversions for flows around a large active region NOAA: 12192 (S12W08) observed on 2014 October 23. 
The flows, plotted as arrows here, are averaged over the near-surface depths of 0 - 3 Mm ($\approx$ 0.99 $R_{\odot}$, {\it left panel}) and 
over 13 - 21 Mm ($\approx$ 0.97 $R_{\odot}$, {\it right panel}). 
The background color image is the LOS magnetic field map of the bipolar spot region.} 
\label{fig4}
\end{figure*}
\begin{figure}[t!]
\includegraphics[width=0.5\textwidth]{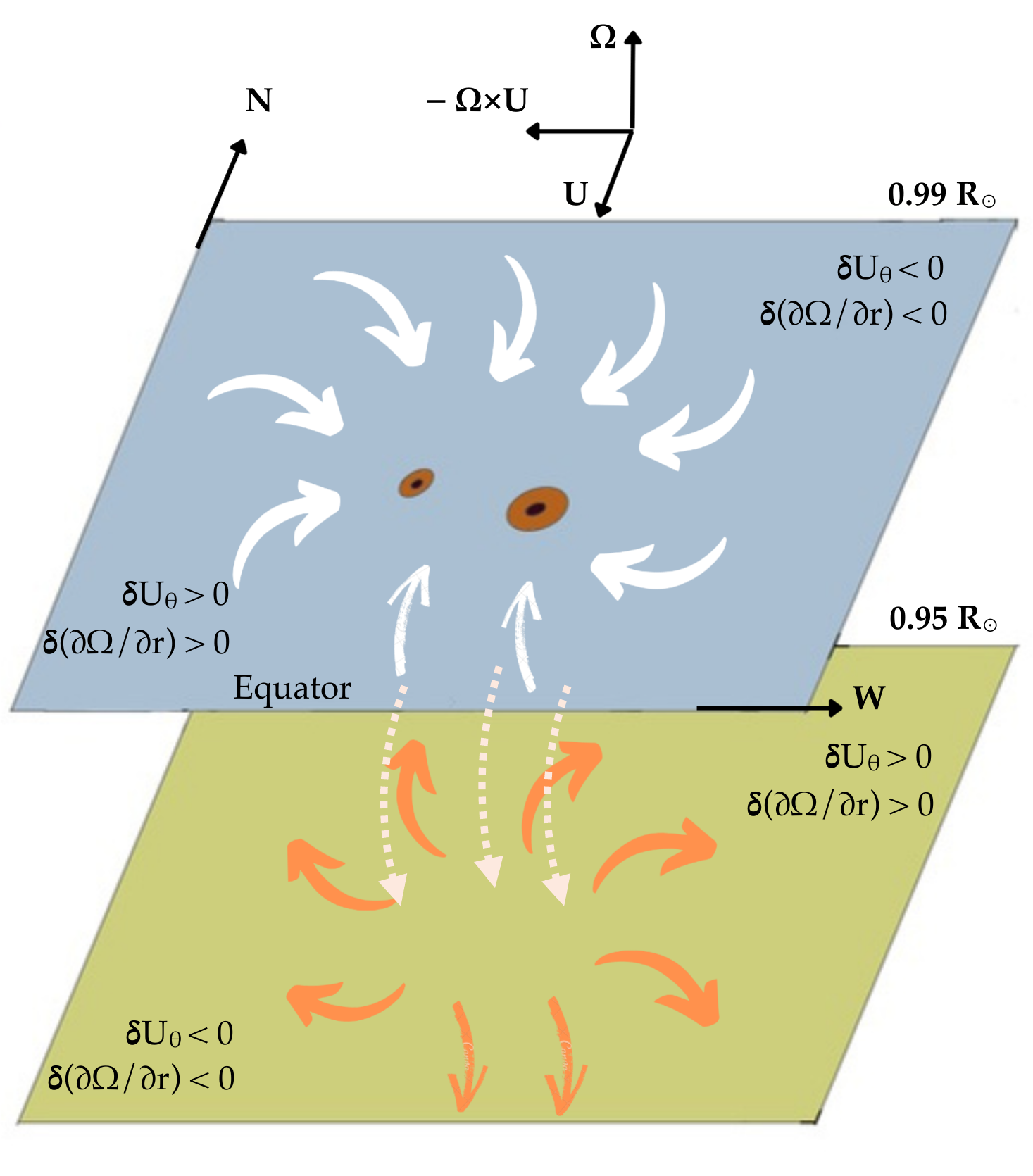}
\caption{Sketch showing the Coriolis force mediated average flow structures around active regions in the northern hemisphere. The labels 
depict the signs of residuals in meridional flows, $\delta U_{\theta}$, and that in resulting residual rotational shear for the two depths,
0.99 and 0.95 $R_{\odot}$, which mark the radial boundaries of the NSSL.} {\em Figure Artwork Credit: Amrita Rajaguru}.
\label{fig5}
\end{figure}

\section{Results} \label{sec: result}

\subsection{Relations Between Residual Rotational Shear and Meridional Flow}
\label{subsec: results_1}

Variations, over time and latitude, in the residual rotational shear,
$\delta(\partial~{\rm{log}}~\Omega/\partial~{\rm{log}}~r)$, and in residual meridional flows, $\delta U_{\theta}$, are
shown in Figure \ref{fig1} for the surface layers (0.99$R_{\odot}$, left panels) and for the bottom layers
(0.95$R_{\odot}$, right panels) of the NSSL. The striking depth and latitudinal dependence of changes
in rotation gradient within the NSSL have already been presented \citep{barekat2016solar,antia2022changes} (except
that results here cover longer into Cycle 25): near 0.95$R_{\odot}$, 
sunspot latitudes coincide with the radial gradient of rotation being less (in magnitude) than average (over the solar cycle), while those
in the near-surface layers (above the depth of 0.98$R_{\odot}$) larger than average. Note that 
radial gradient itself is negative in the NSSL, and hence a negative value for changes in it means it is steeper than
average and vice versa.
The opposite signs of changes at active latitudes at depths 0.99$R_{\odot}$ and 0.95$R_{\odot}$ are clear. 
The sign change happens around the depth of 0.97$R_{\odot}$, and we focus on it later in this Section. Note that the 
global helioseismic measurements are hemisphere symmetric, {\em i.e.,} they are insensitive to hemisphere asymmetric changes. 

The time - latitude profile of $\delta U_{\theta}$ measured using HMI and 
GONG data are shown in the middle and bottom panels of Figure \ref{fig1}. 
Note that, in our sign convention, positive values (red) correspond to poleward and negative (blue) to equatorward flows, 
and hence any cross-equator flow exhibits a discontinuity in colors across the equator in these panels.
As already well established \citep{hindman2009subsurface,2019ApJ...873...94B,2022A&A...664A.189P,mahajan2023removal}, active regions 
drive an inflow towards themselves from the neighbhouring quiet-Sun areas in the near-surface layers. Such flows manifest as negative 
changes on the poleward side and as positive changes on the equatorward side of active belt, in the 
panels for $\delta U_{\theta}$ at 0.99$R_{\odot}$ in Figure \ref{fig1}; the $\delta(\partial~{\rm{log}}~\Omega/\partial~{\rm{log}}~r)$ 
follow suit with a correlated variation (top left panel; see also the sketch in Figure \ref{fig5}). During solar minimum period, 
enhanced poleward meridional flow ($\delta U_{\theta} > 0$) coincides with a less than average residual rotational shear 
($\delta(\partial~{\rm{log}}~\Omega/\partial~{\rm{log}}~r) > 0$). 
These results clearly implicate the E-W (rotational) component of the Coriolis force, $-2{\mathbf\Omega} \times {\delta\mathbf U}$.

Below 0.97$R_{\odot}$, the $\delta U_{\theta}$ reverses sign (0.95$R_{\odot}$ panels in Figure \ref{fig1}). 
Active latitudes are flanked by outflows on either side indicating circulation cells that connect to surface inflows 
({\it cf.} Figures \ref{fig4} and \ref{fig5}). Note that these outflow signals are patchy and we discuss them further below. 
During cycle minimum period, a negative $\delta U_{\theta}$ appears indicating a return flow from about 50° latitude towards the equator,  
possibly fed by the enhanced near-surface poleward flow.
Although a correlated negative $\delta(\partial~{\rm{log}}~\Omega/\partial~{\rm{log}}~r)$ appears at low latitudes,
its latitudinal extent, unlike in the near-surface layers, is narrow and lacks a clear corrspondence with $\delta U_{\theta}$.
We discuss possible physical causes for this again in Section \ref{sec: conclusion}. 
Moreover, the latitude extent of the cycle minimum circulation cell gradually decreases from the higher latitude side and 
becomes the equator-side circulation of the migrating zonal (activity) band of next cycle. 

We point out that the patchiness of $\delta U_{\theta}$ at 0.95$R_{\odot}$ is likely related to the situation of only large active regions
driving significant outflows at depths beneath 10 - 15 Mm \citep{2004SoPh..220..371H}, and also that such signals are significantly above noise 
levels. For example, in the HMI results for solar maximum ({\it cf.} Figure \ref{fig2}), the outflow signals on the poleward side of active latitudes 
(between 20° and 40°) are in the range of 3 to 8 m/s, with estimated errors ranging from 0.3 to 1.1 m/s, and hence signal levels are 3$\sigma$ or higher. 
Here, errors are calculated from inverted flow velocities determined by repeating the inversions 1000 times with travel times randomly perturbed with estimated
errors in observed values \citep{rajaguru2015meridional}.
Further, a significant part of patchiness near the equator is due to cross-equator flows \citep{2022SoPh..297...99K,mahajan2023removal}, 
and we defer a detailed analysis of them to a separate study. We note that a three-year smoothing brings out the correlated  
variations in $\delta U_{\theta}$ and $\delta(\partial~{\rm{log}}~\Omega/\partial~{\rm{log}}~r)$ at 0.95$R_{\odot}$ much clearer (not shown).
GONG and HMI results for $\delta U_{\theta}$ agree well overall for the dominant features discussed above, increasing their reliability. 
However, GONG observations lack sensitivity to flows very near the surface \citep{2020Sci...368.1469G,antia2022changes} and consequently 
suffer noisy fluctuations throughout the near-surface layers, especially at higher latitudes. For all our further presentation in this paper, we use only HMI results.

To understand better the spatial associations between active latitudes and the flows,
we plot latitude profiles of the latter averaged over one year (2014 - 2015) around the solar maximum in Figure \ref{fig2}:
residual meridional flows ($\delta U_{\theta}$, in black) and zonal flows ($\delta U_{\phi}$, in pink)
are shown in the upper panel, and $\delta(\partial~{\rm{log}}~\Omega / \partial~{\rm{log}}~r)$
along with the longitudinally averaged unsigned magnetic field (calculated from HMI LOS magnetograms) in the lower panel;
solid and dashed curves are for depth 0.99$R_{\odot}$ and 0.95$R_{\odot}$, respectively.
We note a few important features in the spatial association between the above quantities:
(i) in the near-surface (0.99 $R_{\odot}$) layers, the latitudinal extent and signs of $\delta(\partial~{\rm{log}}~\Omega / \partial~{\rm{log}}~r)$ 
match with that of $\delta U_{\theta}$, as expected of the action of Coriolis force 
on the latter; near the base of the NSSL (0.95$R_{\odot}$, black dashed curves), however, the former's latitudinal structure coincides with the zonal flow 
($\delta U_{\phi}$) itself, while its sign correlates with that of poleward side $\delta U_{\theta}$. 
Note that the cross-equator flows driven by N-S asymmetry in active regions 
appear as discontinuities at the equator in Figure \ref{fig2} due to our sign convention. (ii) The zonal flows ($\delta U_{\phi}$) 
peak on the equatorward side of active belt \citep{spruit2003origin}, with a slight enhancement on the poleward side (at 0.95$R_{\odot}$)  
because of the outflow ($\delta U_{\theta} > 0$) on that side. (iii) A clear association
between the N-S asymmetry in magnetic field and that in $\delta U_{\theta}$ (upper panel of Figure \ref{fig2}) for both the inflows and
outflows is evident; more interestingly, outflows at 0.95$R_{\odot}$ (dashed curve) show a larger N-S asymmetry in proportion to that in magnetic
field than the inflows near the surface do. The features (i) and (ii) above together show that the horizontal part of flows towards active latitudes
($\delta U_{\theta}$) and the action of Coriolis force on them only minimally alter the zonal flow amplitudes and hence cannot be their cause.

Correlated change of sign over time and depth of residual meridional flows and rotation gradient within the NSSL is the most striking
feature of our results in Figure \ref{fig1}. To bring out clearly such connections between $\delta U_{\theta}$ and 
$\delta(\partial~{\rm{log}}~\Omega / \partial~{\rm{log}}~r)$,
we plot their time - depth profiles in Figure \ref{fig3}. In view of the latitudinal associations brought out above (Figure \ref{fig2}),
it is useful to relate the depth profiles of $\delta U_{\theta}$ over latitudes slightly higher than that of the rotation gradients, and
hence we choose latitudes 20°, 25°, and 30° for the former (top and middle rows of Figure \ref{fig3} for the north and the south, respectively) 
and 10°, 15°, and 20° (bottom row of Figure \ref{fig3}) for the latter. 
During cycle maximum, the prominent equatorward flows (blue) seen from the surface down to 
depths of $\approx$ 0.97$R_{\odot}$, in both hemispheres, are the inflows towards the active latitudes. At depths below 0.97$R_{\odot}$, 
these inflows change direction towards the poles (red), indicating a local circulation connected by downflows
beneath active latitudes. Near the active latitudes (20° and 25°), the strength of this return flow (poleward) away from active 
latitudes, over depths 0.94 - 0.97$R_{\odot}$, is stronger in the southern hemisphere than in the north, clearly correlating 
with the amount of magnetic flux (or the hemispheric asymmetry in the active region magnetic fields). We note that this hemispheric 
difference in peak flows at 0.95$R_{\odot}$ (dashed curve in the upper panel of Figure \ref{fig2}) during solar maximum is about 5 m/s while
the noise levels are in the range of 0.3 to 1.1 m/s. During cycle minimum, the residual flows have the opposite structure over depth. 
The above time - depth profiles of $\delta U_{\theta}$ clearly show a strong
correlation with $\delta(\partial~{\rm{log}}~\Omega / \partial~{\rm{log}}~r)$ shown in the bottom row of Figure \ref{fig3}. 
We point out that, as can be discerned by a close comparison of depth profiles, the change
from inflows to outflows in $\delta U_{\theta}$ happen at a slightly shallower depth than for that in 
$\delta(\partial~{\rm{log}}~\Omega / \partial~{\rm{log}}~r)$. Such a difference is entirely in order if the Coriolis force acting on the former 
is the cause of the changes in the latter, which should occur at the sign change of the depth gradient of the flow rather than that of the flow itself.
It is also interesting to note that the hemispheric change in $\delta U_{\theta}$ over time
occurred earlier in the north, in correlation with hemispheric activity maxima.

\subsection{Flows Around an Active Region}

Among a good number of studies of flows around active regions ({\it e.g.}, \citet{2004SoPh..220..371H,
hindman2009subsurface,2019ApJ...873...94B,2021A&A...652A.148G,2022A&A...664A.189P,mahajan2023removal}), the former two that employed ring diagram
analysis have shown the existence of outflows at deeper layers beneath
large active regions. Here, to ascertain further that the changes we measure in meridional flows in fact relate to the flows around
active regions, especially to depths down to 0.97$R_{\odot}$, we examine 3D (latitude, longitude, depth) local time-distance 
helioseismic inversions for horizontal velocity fields available in the JSOC time-distance (TD) helioseismology pipeline \citep{2012SoPh..275..375Z}. 
We remove the CLS in these full-disk flow maps, stacked from 30$\times$30$^{o}$ tiles, and the large scale time-averaged background flows 
(rotation and meridional flows) following \citet{mahajan2023removal}. We note that these TD pipeline measurements do not mask out
the strong magnetic field umbral pixels and hence are subject to artefacts in flow inferences over the spot area including the moat flow ($\sim$ 50 m/s)
that extend 10 - 20 Mm beyond the penumbra \citep{hindman2009subsurface}; however, we are
concerned with flows surrounding the spots on a larger scale ($\sim$ 100 Mm), where such artefacts are negligible, and moreover we smooth the flow maps
on scales of 20 Mm. 
An example map of flows around a large active region NOAA 12192 (S12W08) observed on 2014 October 23 is shown Figure \ref{fig4}. 
The flows, plotted as arrows here, are averaged over the near-surface depths of 0 - 3 Mm ({\it left panel}) and over 13 - 21 Mm ({\it right panel}).
It is clear that there is a large-scale inflow, starting from around a distance of 100 - 150 Mm away from the spot region, drawing
fluid from the quiet-Sun. 
These converging flows near the surface sink down closer to the spots, and drive an outflow at depths below about 13 Mm (right panel 
of Figure \ref{fig4}). 

On average, the Coriolis force causes cyclonic inflows \citep{hindman2009subsurface,
2019ApJ...873...94B} and anti-cyclonic outflows: a sketch of average flow pattern in the northern hemisphere along with their 
influence on the rotation gradient is shown in Figure \ref{fig5}; note that the active region shown in Figure \ref{fig4} belongs to the southern hemisphere
and hence a flow structure opposite to that shown in Figure \ref{fig5} will apply. We note that flows around
an individual active region will always have deviations over smaller scales and due to complexities in magnetic structure;
an averaging over several active regions is typically needed to bring out the Coriolis force influence on the flows. 
The average structure in Figure \ref{fig5} is consistent with our main inferences on changes
in meridional and zonal flow gradients (Figures \ref{fig1}, \ref{fig2} and \ref{fig3}), which result from the collective (longitudinally
averaged) contributions of all active regions over each measurement period. We however note that further detailed analyses are
needed in light of the most recent findings of \citet{2024ApJ...972..160B} (see Section \ref{sec: conclusion}).

We point out that the transition to outflows for individual active regions happens at 
depths shallower than 0.97$R_{\odot}$, slightly different from the inferences in our global measurements (Figures 
\ref{fig1},\ref{fig2} and \ref{fig3}). This could result from the averaging involved in meridional flow measurements, and could be 
examined in detail in future studies. 

\section{Discussion and Conclusion} 
\label{sec: conclusion} 

In this work, through a joint analysis of radial gradient of rotation and meridional flows from global 
and local helioseismic measurements, respectively, we have studied in detail the solar cycle related  
changes in them within the NSSL. We find that (Figures \ref{fig1} and \ref{fig2}), between 0.97 - 0.99$R_{\odot}$, 
the steeper than average radial gradient of rotation ($\delta(\partial~{\rm{log}}~\Omega / \partial~{\rm{log}}~r) < 0$) 
on the poleward side of active region latitudes is due to inflows ($\delta U_{\theta} < 0$) directed towards them from higher latitudes. 
At depths below 0.97$R_{\odot}$, the smaller than average radial gradient of rotation ($\delta(\partial~{\rm{log}}~\Omega / \partial~{\rm{log}}~r) > 0$) 
is observed to accompany outflows ($\delta U_{\theta} > 0$) away from active latitudes indicating a local circulation connected 
by downflows beneath them. A similar circulation cell on the equatorward side of active regions has a sign-reversed structure of
above residuals in flows and rotation gradient. 

The above connections between residual meridional flows and rotational shear arise from the action of Coriolis force
on active region flows, whose average structure in northen hemisphere is depicted in Figure \ref{fig5}. Longitudinal averaging
of flows around large number of active regions cancels out the inflows in the E-W (rotational) direction \citep{mahajan2023removal}, and
the resulting residual meridional flows take the form of circulation cells on either side of active latitudes as described above.
In the context here, the relevant term in the E-W (or azimuthal) component of Coriolis force is the one involving the meridional 
component of flow: $-2({\mathbf\Omega} \times {\delta\mathbf U})_{\phi} = 2\Omega \sin(\theta)\delta U_{\theta}$, which for equatorward flow 
($\delta U_{\theta} < 0$) is eastward (negative $\phi$ direction) on the Sun and the reverse for poleward flow ($\delta U_{\theta} > 0$), 
$i.e.$, the inflows form a cyclonic circulation because of Coriolis force. Such flows thus tend to slow down the zonal flows on their poleward side
resulting in the steeper radial gradient near the surface, and the opposite result at depths below 0.97$R_{\odot}$ where the
inflows turn into outflows away from active latitudes. Similarly, on the equatorward side of the active belt, the opposite 
influence of flows cause weaker radial gradient (in magnitude) near the surface and vice versa for deeper layers,
but with a reduced effect at low latitudes because of the $\sin(\theta)$ term. Here, if the inflows towards an active region develop 
responding to a horizontal pressure gradient set up by whatever process, say for example by thermal causes, then its balancing with Coriolis force 
will tend to drive the flows geostrophic at low Rossby numbers \citep{spruit2003origin}.
During solar minima, active latitudes which had the near-surface equatorward flows ($\delta U_{\theta} < 0$) are replaced with enhanced
poleward flows (Figures \ref{fig1} and \ref{fig3}), which appear to drive a return flow around 0.95$R_{\odot}$ from about 50° latitude 
towards the equator. Although such solar minimum flows raise the question of their origin, the action of Coriolis force on them is
consistent with the sign change over depth in $\delta(\partial~{\rm{log}}~\Omega/\partial~{\rm{log}}~r)$. 

An important aspect of active region flows has been their cumulative contributions to global scale flows and influence on magnetic flux transport
\citep{2012A&A...548A..57C,2024SoPh..299...42T}.
We note that several studies have examined the level of contribution from active regions to global scale changes in 
meridional flows, with differing conclusions \citep{2008SoPh..252..235G,2010ApJ...720.1030C,2022A&A...664A.189P,mahajan2023removal}.
Our main finding, viz., that the Coriolis force mediation of active region flows is the primary cause of 
correlated variations of $\delta U_{\theta}$ and $\delta(\partial~{\rm{log}}~\Omega / \partial~{\rm{log}}~r)$ (Figures \ref{fig1} - \ref{fig3}),
largely aligns with the case of former changes representing the longitudinally averaged collective contributions of active region 
flows \citep{spruit2003origin,2010ApJ...720.1030C,2019ApJ...873...94B}. The most recent detailed
analyses of \citet{2024ApJ...972..160B}, however, show that active region flows alone cannot explain the $\delta U_{\theta}$. 
Extending our analyses of connections between $\delta U_{\theta}$ and $\delta(\partial~{\rm{log}}~\Omega / \partial~{\rm{log}}~r)$ 
to the level of individual active regions over depths down to the base of the NSSL would be an additional avenue to probe the above issues. 
Hence, we plan on improving the time-distance helioseimic pipeline measurements \citep{2012SoPh..275..375Z} with larger patches 
(than the present 30$\degr \times$ 30$\degr$ tiles) to image the whole NSSL in latitude and longitude.


As noted in Section \ref{subsec: results_1}, the latitudinal extent of negative $\delta(\partial~{\rm{log}}~\Omega/\partial~{\rm{log}}~r)$ 
at 0.95$R_{\odot}$, unlike in the near-surface layers, is narrow and lacks a clear corrspondence with $\delta U_{\theta}$. 
We point out that understanding these features requires consideration of additional physics involving the radial component of circulating flows
\citep{1992ASPC...27..241G,spruit2003origin} acted on by the non-traditional second term in the full expression for the azimuthal (E-W) component of Coriolis 
force: $-2({\mathbf\Omega} \times {\delta\mathbf U})_{\phi} = 2\Omega \sin(\theta)\delta U_{\theta} - 
2\Omega \cos(\theta)\delta U_{r}$. Here, an upflow ($\delta U_{r} > 0$) deflects eastward (on the Sun) opposing the rotation while a downflow deflects
westward enhancing it; since these effects go as $\cos(\theta)$, they are maximum at the equator. Near 0.95$R_{\odot}$, the radial flow 
amplitudes are in the range of 0.5 - 1 m/s \citep{rajaguru2015meridional,2020Sci...368.1469G}\footnote{Mass conservation constraints in meridional 
flow inversions yield $U_{r}(\theta,r)$ profiles too.}, which will experience comparable azimuthal Coriolis acceleration as that by a 
meridional flow of 5 m/s near 10° latitude. Hence, we believe that, while the near-surface dynamics is dominated by Coriolis force on horizontal 
component of meridional flows, the deeper layers' at low latitudes could be influenced significantly by that on the radial flows.

As to the origin of inflows towards active regions, \citet{spruit2003origin} proposed the subsurface temperature
variations caused by the increased radiative losses through the plages and small-scale magnetic fields surrounding sunspots.
However, our results in Figures \ref{fig2} and \ref{fig3} show that, near the active latitudes (20° and 25°) and during solar maximum,
the outflows below 0.97$R_{\odot}$ have a larger hemispheric asymmetry in proportion to that in magnetic field than the inflows do near the surface.
This runs counter to the expected stronger correlation of near-surface cooling driven inflows with magnetic flux. Thus,
our results raise the interesting possibility of large-scale active region flows being driven at a deeper location within the NSSL, similar
to an earlier reported finding based on travel-time signatures alone \citep{beck2002new}. 
Here, we would like to draw attention to the proposal by \citet{1992ASPC...27..241G}, based on the dynamics of 'thermal shadows' 
discussed by \citet{1987ApJ...321..984P}, wherein a broad bundle of toroidal field near the base of the convection zone drives a pair of 
meridional circulation cells, which share a downflow over the central latitude of toroidal band. 
We note that an outflow associated with accumulating active region magnetic field near the base of the NSSL, 
either due to thermal causes or directly due to emergence and horizontal separation, could draw fluid from the overlying layers initiating downflows,
which then drive inflows near the surface forming a circulation cell, similar to the proposal of \citet{1992ASPC...27..241G} for the fields near the tachocline.
We anticipate closer and detailed examinations of our findings in the near future, and that such studies
are crucial to understanding the origin of zonal and meridional flows and their fundamental connections to
solar magnetism.

Irrespective of the basic causes behind, our findings, especially that in Figure \ref{fig3}, show that the
meridional and zonal flows have correlated variations over solar cycle time scales, although
the amplitudes of latter change only marginally (cf. Figure \ref{fig2}). We conclude that the Coriolis force on the residual 
meridional flows causes the change in the radial gradient of zonal flows within the NSSL, while the driving of zonal flow 
itself is likely of deeper origin. We point out, in closing, that the close agreements
and correlated variations between the depth profiles of changes in rotation gradient and in meridional flows measured in this work
here from the quite different global and local helioseismic techniques, respectively, provide an important validation for the measurement
procedures, especially for the latter.

\section{Acknowledgements}
The HMI data used are courtesy of NASA/SDO and the HMI science teams. Data preparation and processing have utilised the Data Record 
Management System (DRMS) software at the Joint Science Operations Center (JSOC) for NASA/SDO at Stanford University. 
We acknowledge that the global helioseismic inversions of solar interior rotation used in this work are from the published results 
of H.M. Antia, and we thank him for sharing them. We also thank the two anonymous reviewers, whose extensive, critical reviews and
suggestions contributed substantially to the interpretation and presentation of results in this paper. We accord special credits to one of
the reviewers for suggestions on Figure 5.
The GONG data used are obtained by the NSO Integrated Synoptic Program, managed by the National Solar Observatory, 
which is operated by the Association of Universities for Research in Astronomy (AURA), Inc. under a cooperative 
agreement with the National Science Foundation and with contribution from the National Oceanic and Atmospheric Administration.
This work has received funding from the NASA DRIVE Science Center COFFIES Phase II CAN 80NSSC22M0162
to Stanford University. S.K. was partly supported by NASA grants 80NSSC21K0735, 80NSSC19K0261 and 80NSSC20K0194 to NSO.
Data intensive computations in this work have utilised the High-Performance Computing
facility at the Indian Institute of Astrophysics. A.S. is supported by INSPIRE Fellowship from the
Department of Science and Technology (DST), Government of India. S.P.R. acknowledges support
from the Science and Engineering Research Board (SERB, Government of India) grant CRG/2019/003786.

\bibliography{msbib}{}
\bibliographystyle{aasjournal}
\end{document}